\documentstyle[multicol,psfig,aps,prl]{revtex}
\draft

\newcommand{\ba}{\begin{eqnarray}}                                             
\newcommand{\ea}{\end{eqnarray}}

\newcommand{\w}{$\omega$}
\newcommand{\la}{\langle}
\newcommand{\ra}{\rangle}

\newcommand{\xy} {$|\chi^{(1)}_{xy}|$}
\newcommand{\zz}{$|\chi^{(1)}_{zz}|$}
\newcommand{\xzz}{$|\chi^{(2)}_{xzz}(\omega,t)|$~}
\newcommand{\bml}{\begin{multicols}{2}}
\newcommand{\eml}{\end{multicols}}

\newcommand{\zzz}{$|\chi^{(2)}_{zzz}(\omega,t)|$~}

\newcommand{\sz}{$|\hat{S}_z(t)|$~}
\newcommand{\nn}{$|\hat{N}(t)|$~}

\newcommand{\be}{\begin{equation}}
\newcommand{\ee}{\end{equation}}  
\begin{document}
\title{Femtosecond electron and
spin dynamics probed by nonlinear optics} 
\author{G. P. Zhang$^*$  and W. H\"ubner}
\address{Max-Planck-Institut f\"ur Mikrostrukturphysik, Weinberg 2,
D-06120 Halle, Germany\\
$^*$
e-mail: zhang@mpi-halle.mpg.de}

\date{\today}

\maketitle

\begin{abstract}
A theoretical  calculation is performed  for  the ultrafast spin  dynamics  in
 nickel using  an exact  diagonalization method.    The present theory  mainly
 focuses on a situation where the intrinsic charge and spin dynamics is probed
 by the nonlinear (magneto-)optical responses  on
 the femtosecond time scale, i.e. optical second harmonic
 generation (SHG) and the nonlinear magneto-optical Kerr effect (NOLIMOKE).
It is  found that the  ultrafast charge and spin
 dynamics are  observable on  the time scale   of 10 fs.  The  charge dynamics
 proceeds ahead of the spin dynamics, which indicates the  existence of a spin
 memory  time.  The  fast  decay results from  the  loss of   coherence in the
 initial excited state.      Both  the material specific    and   experimental
 parameters affect  the  dynamics.   We find  that  the  increase of  exchange
 interaction  mainly accelerates  the  spin  dynamics  rather than the  charge
 dynamics.   A reduction  of   the  hopping integrals,    such as present   at
 interfaces, slows down the spin dynamics significantly.  Besides, it is found
 that  a spectrally broad excitation yields  the intrinsic speed limit of the
 charge (SHG) and spin dynamics (NOLIMOKE) while a narrower width prolongs the
 dynamics.  This magnetic interface dynamics  then should  become accessible
 to state of art time resolved nonlinear-optical experiments.

\end{abstract}

\bml
\newpage

\section{Introduction}

Recently ultrafast spin dynamics in ferromagnetic metals has attracted a great
deal of  attention due to its  possible applications, such as the ultrafast
magnetic gates.  
The experimental observation  was   that,  upon  the  excitation   of
a femtosecond laser pulse, a sharp decrease of the magnetization occurs on a
time scale of 100 fs\cite{eric,hohl,aesch,scholl}, 
which is  far beyond the characteristic
time  scale  of    spin-lattice   interaction.   Similar results   have   been
independently found in  pump    and probe linear    magneto-optics,  nonlinear
magneto-optics  and two-photon-photoemission.  However  the  interpretation of
this  behavior  has not been given  on  the same  footing and remains somewhat
speculative. To reconcile these intriguing 
results  from  different experimental  processes  such as linear and nonlinear
optics is rather crucial and  is one of the  goals of the present theoretical
study.  Moreover  the demagnetization is   very  similar to  the  conventional
one\cite{hohl}.

In the SHG experiment \cite{hohl}, the $M(T)$ curve is established
after the  electron thermalization is finished  while the electron and lattice
have not  reached a  common equilibrium yet,  which  indicates  a  purely electronic
feature  of  the ultrafast  spin   dynamics.  Traditionally  the  spin-lattice
coupling sets
the speed  limit of the demagnetization process,  typically on 100 ps\cite{meier},
but here this is clearly not the case.  Our previous theoretical studies
clearly demonstrated that  the dephasing of initial  states  is the origin  of
the spin dynamics  on the femtosecond
 time  scale \cite{hz1,hz2}, as probed by
transient reflectivities and linear magneto-optics. We found that the 
intrinsic speed limit is about 10 fs.  This mechanism 
is a pure quantum effect, resulting
from  the interplay between band structure  and  electron correlation. In this
paper, we focus on the {\em nonlinear} 
optical response of spin dynamics. We take a Ni monolayer as an example. 
 
This paper is arranged as follows. In Section II, we discuss our theoretical
scheme while the main results are given in Section III. Finally,
the conclusions are given in Section IV.

\section{Theoretical scheme}

It is well-established that 
in  ferromagnetic transition metals, in particular in  Ni, 
the electron correlation plays an important
role even in the ground state 
and  possesses a significant impact on the 
excited states, as  evidenced e.g. by the famous photoemission  satellite
structure\cite{fulde}. This becomes especially true in the
nonlinear optical process on the ultrafast time scale where highly excited
states are frequently involved. 
Therefore, we employ an exact-diagonalization
method  which explicitly avoids a perturbative treatment of electron
correlation.
Within our scheme,  one
 does not need to introduce 
any damping term to obtain a correct dephasing
time. Our previous results showed that the typical time scale is basically set
by the dispersion of the bands and  the strength of electron correlation.  
We begin with a generic Hamiltonian

\ba
H=\sum_{i,j,k,l,\sigma,\sigma',\sigma'',\sigma'''}
U_{i\sigma,j\sigma',l\sigma''',k\sigma''}
c_{i\sigma}^{\dag}{c}_{j\sigma'}^{\dag}c_{k\sigma''}{c}_{l\sigma'''}\nonumber 
\\
+\sum_{\nu,\sigma,K}{\cal E}_{\nu}(K)n_{\nu\sigma}(K)+H_{SO} 
\ea

\noindent  where $U_{i\sigma,j\sigma',l\sigma''',k\sigma''}$ is the on-site  
electron
interaction, which can be described in full generality by the three parameters
Coulomb repulsion  $U$, exchange interaction  $J$, and the exchange anisotropy
$\Delta J$ \cite{hf}. The set of parameters used for Ni is 
given  in  Ref.  \cite{wh}.  $c_{i\sigma}^{\dag}$
(${c}_{i\sigma}$)  are   the usual creation    (annihilation) operators in the
orbital   $i$    with spin  $\sigma$   ($\sigma=\uparrow,\downarrow$).  ${\cal
E}_{\nu}(K)$  is the single-particle energy  spectrum for band $\nu$ of the 
nickel monolayer.  
$n_{\nu\sigma}(K)$  is the  particle number  operator in momentum
space.  $H_{SO}$  is  the spin  orbit  coupling.  Since this is  
a typical many-body 
problem, one cannot solve it   without simplification. In order  to
obtain a tractable model,  we first build  a  two-hole basis set.  Within this
basis set,  for each Ni atom,  the dimension of the  Hilbert space is  66. The
matrix  elements of  electron  correlation  for each    atom can be   obtained
analytically. For each $K$ point, the electron correlation is embedded in the
crystal  field as given by the  bandstructure. This treatment of correlations
is   analogous to  a 
frequency dependent  self-energy   correction.  Within this
simplification, we are  able to exactly   diagonalize the Hamiltonian  for each $K$
point  explicitly.

In order to characterize the spin and charge dynamics clearly, we calculate
both 
these intrinsic quantities:
$S_z(t)=\langle\Psi(0)|\hat{S}_z|\Psi(t)\rangle$ and
$N(t)=\langle\Psi(0)|\hat{N}|\Psi(t)\rangle$, 
and the nonlinear (magneto-)optical susceptibilities  $\chi^{(2)}_{xzz}$ and
$\chi^{(2)}_{zzz}$. Here
$\hat{S_z}=\frac{1}{2}(\hat{n}_{\uparrow}-n_{\downarrow})$,
$\hat{N}=(\hat{n}_{\uparrow}+n_{\downarrow})$, which are directly related to
the observable NOLIMOKE and SHG yields, respectively. 
Since \xzz and \sz mainly reflect 
the spin response  while \zzz and \nn reflect the charge response, they will
be used as indicators to evaluate spin and charge evolutions, respectively.
We find:
{\small 
\ba
\chi^{(2)}_{xzz}(\omega,t)&=&\sum_{k,l,l',l''}
(
\frac{p(E_{kl''},t)-p(E_{kl'},t)}
{E_{kl''}-E_{kl'}-\omega+i\eta}\nonumber\\
&-&
\frac{p(E_{kl'},t)-p(E_{kl},t)}
{E_{kl'}-E_{kl}-\omega+i\eta}
)
/({E_{kl''}-E_{kl}-2\omega+i2\eta})\nonumber \\
&\times& (\la kl|\hat{S}_z|kl \ra +
\la kl'|\hat{S}_z|kl'\ra +\la kl''|\hat{S}_z|kl''\ra-3/2)\nonumber\\
\ea

\ba
\chi^{(2)}_{zzz}(\omega,t)&=&\sum_{k,l,l',l''}(
%\frac{
\frac{p(E_{kl''},t)-p(E_{kl}',t)}
{E_{kl''}-E_{kl'}-\omega+i\eta}\nonumber\\
&-&
\frac{p(E_{kl'},t)-p(E_{kl},t)}
{E_{kl'}-E_{kl}-\omega+i\eta})
%}
/{(E_{kl''}-E_{kl}-2\omega+i2\eta)}\nonumber\\
\ea

}
\noindent where $|kl>$ is the eigenstate with the eigenvalue $E_{kl}$;
$p(E_{kl},t)=<\Psi(t)|kl>$.

\section{results}

Before we come to our main results, we would like to demonstrate that our
Hamiltonian can reasonably describe some basic experimental results. 
It has been well-established that a prerequisite to acquire a ferromagnetic
ground state is a nonzero Coulomb interaction $U$ and exchange interaction
$J$. We can simply check this by setting both $U$ and $J$ to zero. Doing so,
we find that the ground state is a singlet, i. e. paramagnetic state, which
contradicts the ferromagnetic nature of nickel. This proves the importance of
$U$ and $J$. Once we use the generic
sets of $U$ and $J$ of nickel, 
we obtain a triplet as its ground state, from which we
can calculate the magnetic moment, 0.88 $\mu_B$. This magnetic
moment is larger than that in  the bulk material, which is consistent with the
experimental observation. In this case, also the satellite structure 
well-known from
photoemission experiments appears in the spectrum quite naturally.

In Fig. 1, we firstly show the effect of exchange coupling $J$ on the \xzz and \zzz as
a function of time $t$. The probe frequency \w~ is  fixed at 2 eV.
The initial state is prepared to be 2 eV above the ground
state with a Gaussian broadening as large as 20 $\!\!\!\!\!\!\!\!$$
\left . \right \lceil$ eV, which opens almost all the
possible decay channels. Such a large distribution width corresponds to a
very large laser spectral width.
As previously explained, this choice aims at revealing the 
real {\em intrinsic}  speed limit of the spin dynamics in our system, which is
then not delayed by experimental constraints. In 
Figs. 1(a) and 1(b), the 
generic parameters of Ni monolayers are used. There are several
interesting features that should be mentioned. One notices that in Fig. 1(a)
\xzz first comes up very 
quickly and reaches its maximum at about 2-3
fs. Then \xzz undergoes a sharply decreasing envelope
 and oscillates with a very short
period. The dynamics of 
 \xzz settles down 
around 10 fs (decay to 1/e of maximum), which
indicates the complete dephasing. As
aforementioned, \xzz signifies the spin relaxation process. Thus we estimate
that the spin relaxation time is about 10 fs, which is consistent
with our previous results for time-resolved linear (magneto-)optics 
 based on \xy~ and \zz. 
For the charge dynamics, we see a 
different scenario. In Fig. 1(b), \zzz is plotted as a function of time
$t$. One sees that \xzz nearly spends a similar time to reach its maximum as
\xzz does, but the subsequent decay occurs more sharply and strongly. Around
5 fs, the value of \zzz is already close to the equilibrium data. This means
that the dephasing already becomes strong for the charge dynamics before the
spin dynamics. If one compares
Fig. 1(a) with 1(b), one sees a clear difference between spin and charge
dynamics. Basically the spin dynamics lasts about twice as long as
the charge
dynamics. This has an important consequence as it demonstrates the spin memory
effect: though the charge dynamics finishes, the spin dynamics is still
alive, which is crucial for future applications. The main difference of the
time resolved nonlinear response compared

\hspace{0.9cm}\psfig{figure=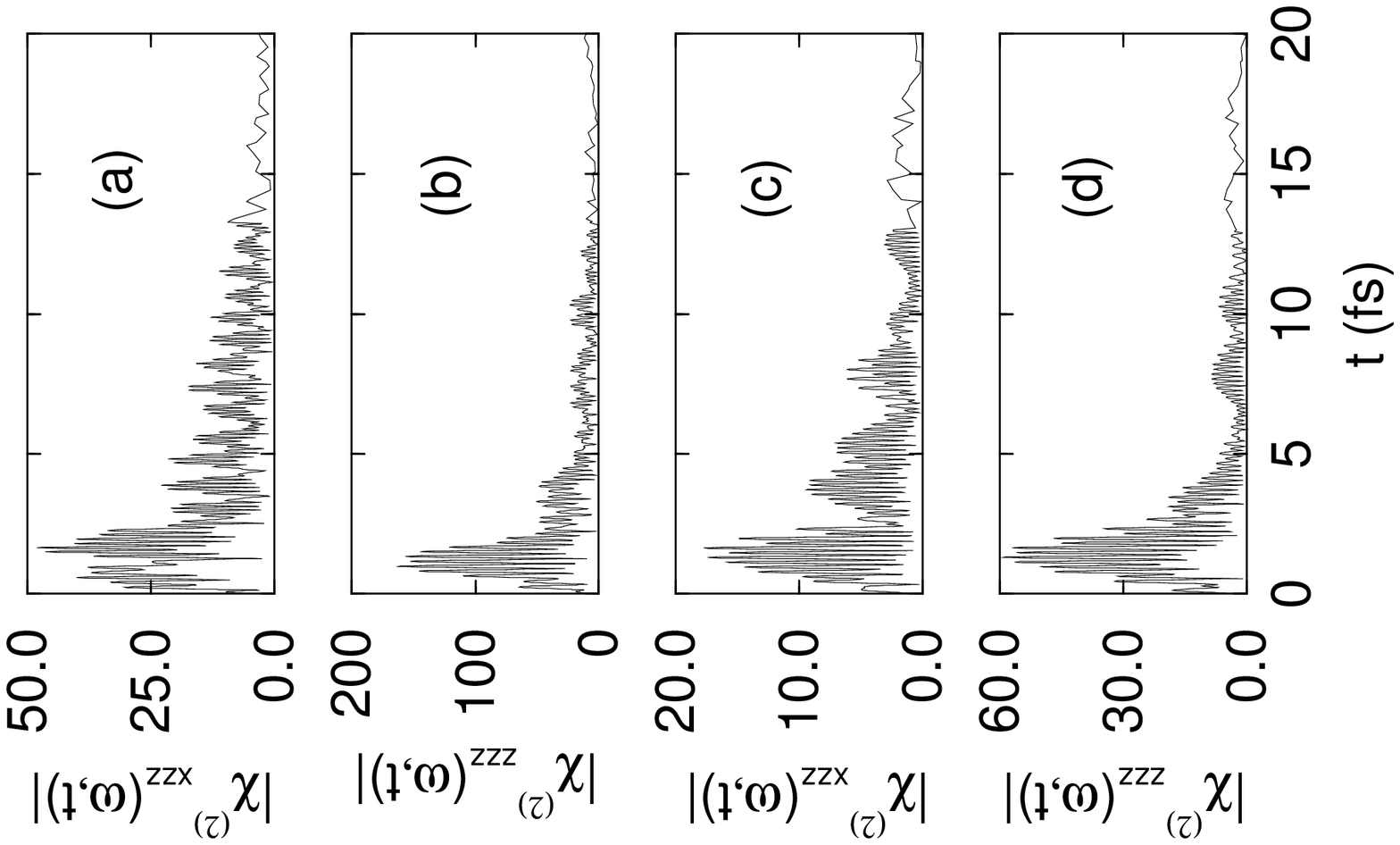,width=6cm,angle=270}
\begin{figure}
\caption{ 
The femtosecond time evolution of the
nonlinear magneto-optical and 
optical responses, \xzz and \zzz, respectively.
Here the initial excited state is prepared 2 eV above the ground
state  with a Gaussian broadening as wide as 20 eV. In (a)  and (b), a set of 
generic parameters of nickel is used; in (c) and (d), the exchange interaction $J$ is
reduced 
to $J_0/10$ while the rest of the parameters is kept 
unchanged. The spin dynamics is delayed
with respect to charge dynamics, which indicates the existence of a spin memory
effect.  Additional `bunching' occurs, which has not been found for the linear
pump-probe responses of Ni.
\xzz oscillates with a larger period.
}
\end{figure}

\noindent to the linear one, which is
particularly evident for the magnetic dynamics, consists in an additional
`bunching' of the structures resulting from the simultaneous presence of \w~
and 2\w~ resonances in Eqs. (2) and (3). 

In order to get a handle on the microscopic origin of the  observed magnetic
dynamics, 
we try to investigate the effect of the  
on-site exchange coupling $J$. We reduce $J$
to $J_0/10$. The  corresponding time-dependences of
 \xzz and \zzz are shown in Figs. 1(c) and 
 1(d),
respectively. It can be seen that \xzz first 
comes up within 2 fs. After that 
a recurrence appears  with a rather large amplitude. Compared with Fig. 1(a), 
\xzz oscillates with a longer `bunching' 
period and the loss of coherence is weaker. We
estimate that the relaxation time is about 10 fs but the period is nearly
twice as long as that in Fig. 1(a).   This demonstrates that
the decrease of exchange interaction prolongs the period of oscillations. 
For the
charge dynamics, the change going from $J_0$ to $J_0/10$ 
is relatively small. This can be seen from
Fig. 1(d) where we plot \zzz as a function of time $t$. Comparing Figs. 1(b)
and 1(d), one finds that the overall variation of \zzz with time is
nearly identical. This is understandable since the 
exchange interaction acts more
directly on the spin degree of freedom
 by changing the  spin dependence of the electronic many-body states
microscopically. Consequently, the spin dynamics will be  affected more
strongly than the charge dynamics.
However electrons even with  different
spin-orientations play a similar role for the charge dynamics. 
Thus, the charge dynamics is basically independent of
 the spin state. That is why
the exchange interaction does not 
\noindent affect the charge dynamics significantly.

Next, we wish to gain some physical insights into the effects of the
experimental constraints, such as the spectral width of the excited state
distribution, on the
spin 
 dynamics. In Fig. 2, we
perform  a detailed comparison between those experimental observables and
intrinsic quantities (i.e., \sz, \nn). The initial excited state is prepared 2
eV above the ground state with a Gaussian broadening of now only 0.2 eV, which
simulates a narrow

\hspace{-0.5cm}\psfig{figure=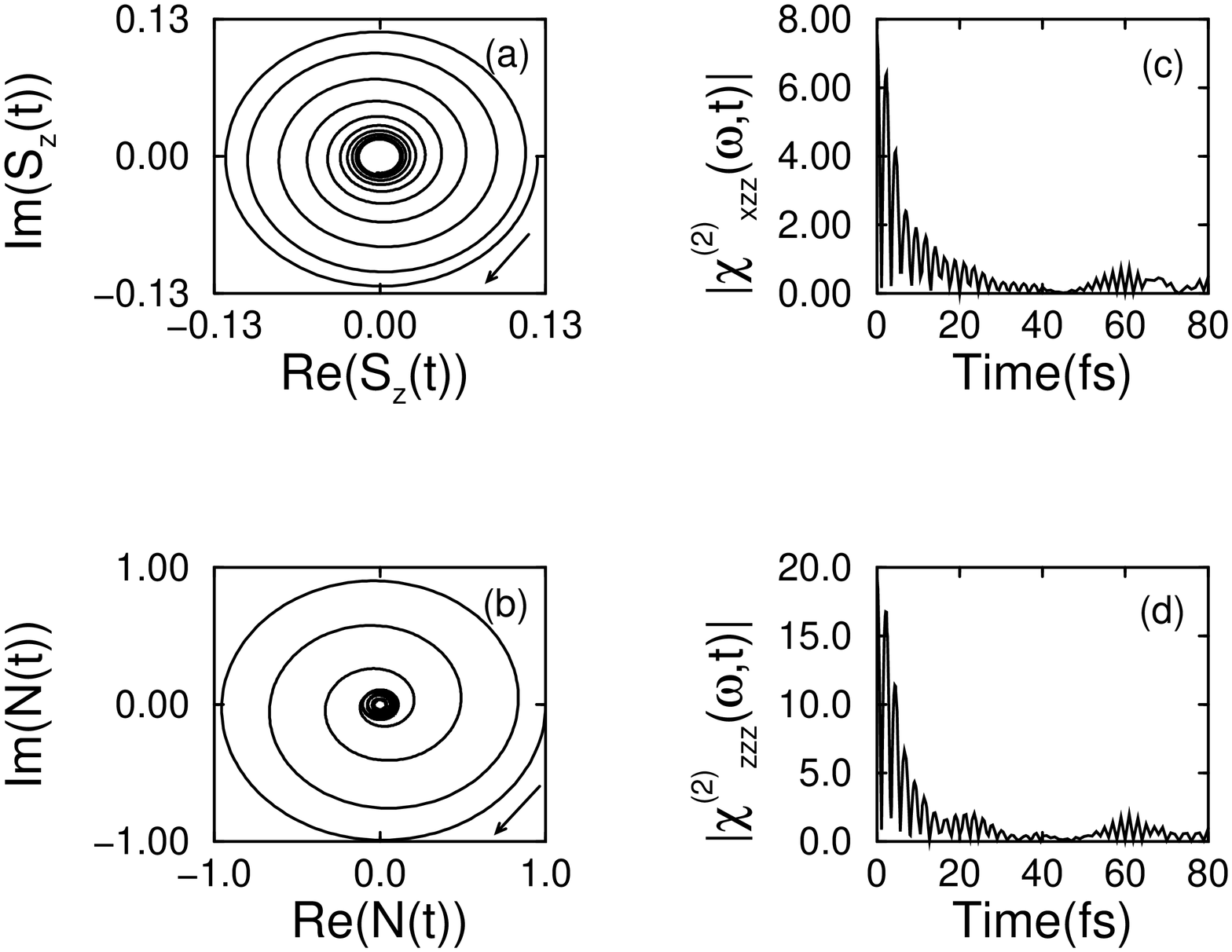,width=8.5cm,angle=270}
\begin{figure}
\caption{  Femtosecond time evolution of the intrinsic spin
(a) and charge (b) dynamics, which is not directly observable, in comparison
with the magneto-optical (c) and optical (d) response functions.
\sz and \nn, 
(a) and (b) show the clearly different behavior between spin
and charge dynamics. Here the initial state is also prepared to be 2 eV above
the ground state, but the spectral width is 0.2 eV. The other 
parameters are taken
as the  generic parameters of Ni. The time interval is [0,40 fs] for (a) and
(b).  
The calculated intrinsic ((a), (b)) and experimentally accessible  ((c), (d)) quantities show
a strong dependence on the laser spectral width (corresponding to the initial
excited state width). Comparing with Fig. 1, we notice that for the same set
of generic  parameters, the decrease of the laser spectral width yields a
slower decay of the spin and charge response.
}
\end{figure}

\noindent  laser spectral width, in contrast to the previously used
width of 20 eV. This allows us to see the effect of the
 laser spectral width clearly.
All the parameters for the Hamiltonian 
are the generic set of parameters of Ni.
In Figs. 2(a) and 2(b),
we show the results up to 40 fs. 
The abscissa and ordinate in Figs. 2(a) and 2(b) denote
the real and imaginary parts of the intrinsic spin and charge dynamics, 
${S}_z(t)$ and $N(t)$, respectively. Note that these quantities cannot
directly be observed and are only theoretically accessible.
The  arrows  refer to the  time  direction and the
centers are the  final positions of $S_z(t)$ and  $N(t)$. One notices that the
spin dynamics needs about six cycles to reach its final value while for the 
 charge
dynamics only three cycles are needed. This demonstrates again that the spin
dynamics is delayed with respect to the 
charge dynamics. We find that comparing
those intrinsic quantities one can see a clear
 difference between spin and charge responses, which is not blurred by the
details of the experimental conditions. 
Nevertheless,
from our results, one  can still 
identify the differences also in the nonlinear optical and magneto-optical responses. We show the results in Figs. 2(c) and 2(d). In
Fig. 2(c), \xzz is plotted as a function of time up to 80 fs. One may notice
that a decay occurs within 30-40 fs,
which is consistent with our previous linear results obtained from \xy~
and \zz\cite{hz1}.  SHG probes more bands, which causes more `beating' and
`bunching'. This might then result in an effectively slower response than that
seen in linear pump-probe experiments, in
addition to the slowing down which results from the narrower bandwidth at
interfaces. 
For \zzz we present the results in Fig. 2(d). Comparing
with \xzz
in Fig. 2(c), \zzz drops even more sharply. One can see a clear drop 
 at 18 fs, which sets its relaxation time for charge dynamics. A delay about
10 fs between spin and charge responses is found for this specific set of
parameters. Comparing Figs. 1(a) and
 2(c), one can immediately see that a narrow
initial state distribution width prolongs the relaxation process. In Fig. 1,
one knows that the relaxation process basically finishes within 10 fs, but
here the relaxation time is around 30-40 fs. 

Finally, as  our previous  studies   already showed \cite{hz1,hz2}  for  the
linear pump-probe calculations, the    band structure will also   influence the
relaxation process. Its  effect is actually  very significant.  Our results on
\xy~  and \zz~ already  showed   that the hopping   integrals  can 
modify  the
relaxation   process strongly. Analogously,  this   will  be reflected in   the
nonlinear optical responses \xzz and \zzz.  In order to investigate the effect
of  the band structure,  we reduce the  hopping integrals to  one-tenth of the
original nickel  hopping  integrals  
 while keeping  the   rest of  parameters
unchanged. Here,
 the initial excited state is also prepared 2 eV above the ground
state  with   a Gaussian broadening  of  20  eV.  The   results  are  shown~in

~~~\psfig{figure=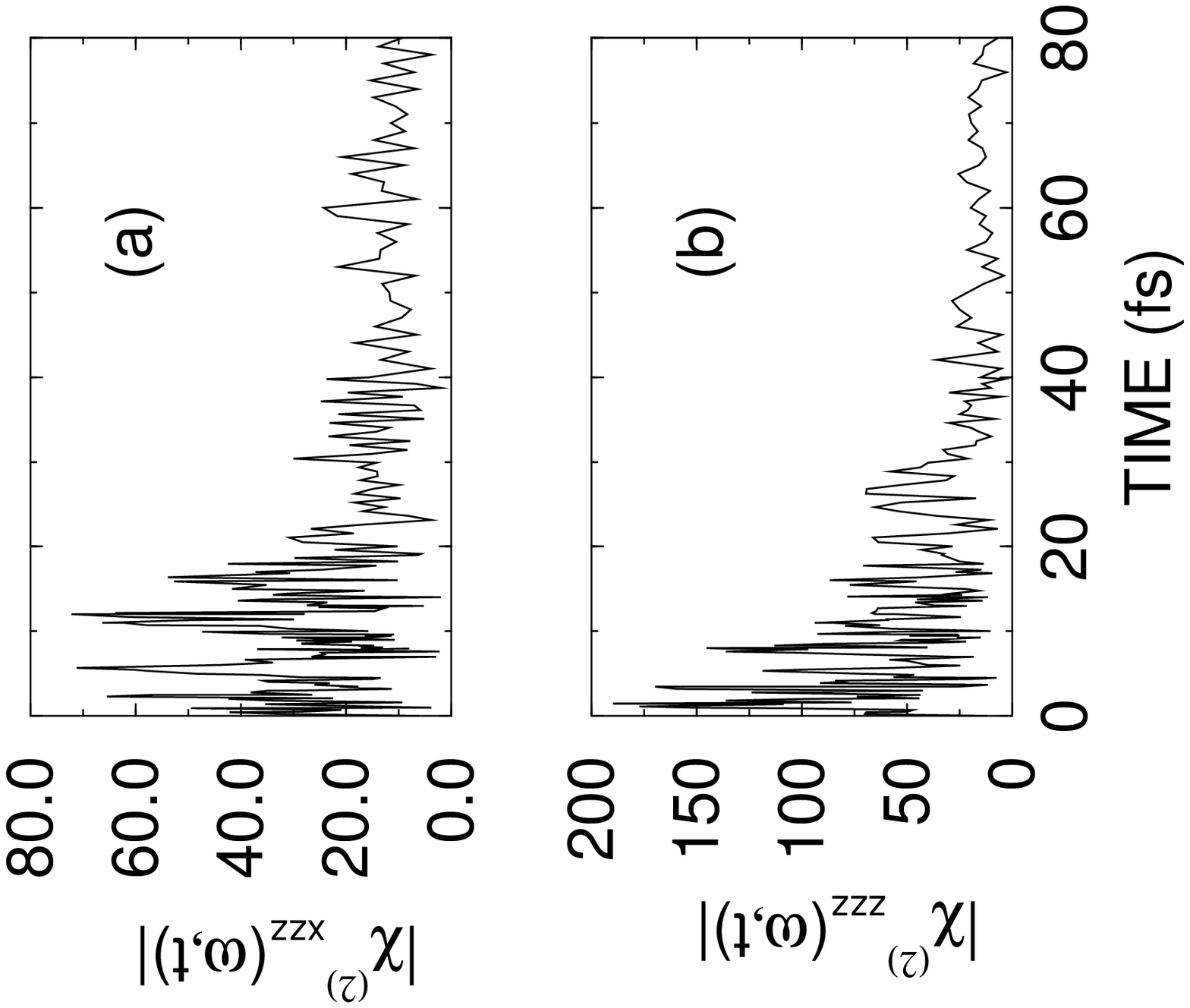,width=7cm,angle=270}
\begin{figure}
\caption{
The femtosecond time evolution of the
nonlinear magneto-optical and 
optical responses, \xzz and \zzz, respectively.
The same initial condition has been used as in Fig. 1 except for the
hopping integrals which are taken to be   one-tenth of the original ones.  
It turns out that 
the change of band structure has a significant impact on spin  and charge 
dynamics. It slows down the dynamics by modifying the envelopes of the response
functions \xzz and \zzz. 
} 
\end{figure}

\noindent 
Fig. 3. One finds  that the change of  both \xzz and   
\zzz with time  is very
different from  the  previous ones.  From  Fig.  3(a),  one notices
 that  \xzz
increases sharply within  5 fs. The strong  oscillation lasts about 20 fs,
where no clear decay can be seen. The outline of \xzz ranging  from 0 fs to 20
fs  forms a broad   peak.  After 20 fs,   dephasing  occurs, but  the envelope
 of \xzz decays only slowly. 
Comparing Figs. 1(a) and 1(b) with Figs. 3(a) and 3(b), respectively, one sees
that the reduction  of the 
hopping integrals 
slows down  both the spin and  charge
dynamics considerably. The  envelope of \xzz now decays  on the time  scale of
30-40 fs.  The appreciable  fast oscillation in  the  long-time tail  survives
beyond   80 fs.  Comparing Figs.  3(a)  and 3(b)   with  Figs. 2(c)  and 2(d),
respectively, indicates that the 
elementary oscillations are  more rapid. Thus from
the comparison of all  the three figures,   we conclude that the reduction  of
the excited  states distribution  width  affects the  
 elementary oscillation more
strongly than the envelope of the spin  and charge responses while decreasing
the  hopping integrals does the opposite  and mainly slows  down the response
envelope of \xzz and \zzz. It is remarkable that in  all cases the spin memory
effect persists. In addition, we would like to point out that the change of
band structure imposed by spin-orbit coupling as well affects the spin
dynamics observed in nonlinear magneto-optical experiments. For linear
time-resolved magneto-optical experiments, this has been demonstrated
theoretically (see Ref. \cite{hz1,hz2}). We found that spin-orbit coupling had
to be raised to an unphysical large value of 1 eV (the generic value of
spin-orbit coupling in nickel is 0.07 eV) in order to be relevant for the spin
dynamics. Thus we do not expect any major influence from spin-orbit coupling
on the 10 fs spin dynamics of transition metals in the valence band.

\section{conclusions}

In conclusion, we performed an exact-diagonalization scheme to study the spin
dynamics in ferromagnetic nickel on the femtosecond time scale. 
This study is exclusively  devoted to the
nonlinear optical responses. At first, we checked that 
our Hamiltonian can give a correct ferromagnetic ground state and a correct
order of magnitude of the 
magnetic moment, which is larger than its value in bulk materials,
as is physically expected. 
 For a generic set of Ni
parameters, the intrinsic speed limit of spin dynamics is about 10 fs, which is
on a similar time scale as revealed by \xy~ although additional `bunching'
structure appears for the nonlinear time-resolved magneto-optical response. 
 Our theory clearly yields the
memory effect of the spin dynamics also in nonlinear optics. 
It is found that the spin dynamics is delayed
with respect to the 
charge dynamics. The spin dynamics survives even when the charge
dynamics ceases. This is very important for future applications, such as 
ultrafast magnetic gates. 
We also examined the effect of exchange interaction. 
It is found that 
a decrease of the  exchange interaction prolongs the relaxation
process. Especially we noticed that for a smaller exchange interaction $J$ 
the spin response `bunches'  with longer periods. This again confirms our
earlier results that the observed spin dynamics results from the spin dependent
($J$-dependent) dephasing in the excited many-body states. In that sense, SHG
and NOLIMOKE reveal the same fundamental physics as time-resolved linear
(magneto)-optics. 
In order to reveal more
insight into the spin dynamics, we also calculated some intrinsic
quantities, from which we see a clearer difference between spin and
charge dynamics. In addition, we studied the laser width effect on the 
spin
dynamics. It shows that a narrow spectral width clearly slows down the
dynamics. This is understandable as, for a smaller spectral width, the number
of  populated states is smaller. The dissipation becomes weaker and eventually
the spin dynamics is prolonged significantly.  Finally we studied the effect
of the band structure on the 
 spin dynamics. We found that the reduction of the hopping integrals slows down
both the spin and charge dynamics by modifying the envelope of the response
functions \xzz and \zzz.
This is different from the effect of a
narrower  distribution width, where one basically prolongs the oscillation
periods.  
It is important to note that materials with 
a small hopping  
integral correspond to  nanostructure materials, clusters, quantum well
states,  magnetic insulators
(such as oxides), defects, and
 impurities. Thus it is expected that with the presence of those nanostructure
materials, the observed dynamics will be significantly slower. Naturally, 
this is very meaningful for the design of materials for ultrafast magnetic
devices. Finally and mostly importantly, the processes which we investigated
all exhibit the spin memory effect, a crucial property for future applications.

%%%%%%%%%%%%%%%%%%%%%%%%%%%%%%%%%%%%%%%%%%%%%%%%%%%%%%%%%%%%%%%%%%%%%
%%%%%%%%%%%%%%%%%%%% F I G U R E 1 %%%%%%%%%%%%%%%%%%%%%%%%%%%%%%%%%%
%%%%%%%%%%%%%%%%%%%%%%%%%%%%%%%%%%%%%%%%%%%%%%%%%%%%%%%%%%%%%%%%%%%%%

%\vspace{3cm}

%\centerline{FIGURE 1, {\it Zhang and H\"ubner}}

%%%%%%%%%%%%%%%%%%%%%%%%%%%%%%%%%%%%%%%%%%%%%%%%%%%%%%%%%%%%%%%%%%%%%
%%%%%%%%%%%%%%%%%%%% F I G U R E 2 %%%%%%%%%%%%%%%%%%%%%%%%%%%%%%%%%%
%%%%%%%%%%%%%%%%%%%%%%%%%%%%%%%%%%%%%%%%%%%%%%%%%%%%%%%%%%%%%%%%%%%%%

\vspace{3cm}

%\centerline{FIGURE 2, {\it Zhang and H\"ubner}}

%%%%%%%%%%%%%%%%%%%%%%%%%%%%%%%%%%%%%%%%%%%%%%%%%%%%%%%%%%%%%%%%%%%%%
%%%%%%%%%%%%%%%%%%%% F I G U R E 3 %%%%%%%%%%%%%%%%%%%%%%%%%%%%%%%%%%
%%%%%%%%%%%%%%%%%%%%%%%%%%%%%%%%%%%%%%%%%%%%%%%%%%%%%%%%%%%%%%%%%%%%%

%\vspace{3cm}

%\centerline{FIGURE 3, {\it Zhang and H\"ubner}}

%%%%%%%%%%%%%%%%%%%%%%%%%%%%%%%%%%%%%%%%%%%%%%%%%%%%%%%%%%%%%%%%%%%%%
%%%%%%%%%%%%%%%% E N D %%%%%% O F %%%%% F I G U R E S %%%%%%%%%%%%%%%
%%%%%%%%%%%%%%%%%%%%%%%%%%%%%%%%%%%%%%%%%%%%%%%%%%%%%%%%%%%%%%%%%%%%%
\end{multicols}


\begin{references}
\bibitem{eric}E. Beaurepaire, J. -C. Merle, A. Daunois, and J. Y. Bigot,
Phys. Rev. Lett. {\bf 76}, 4250 (1996).
\bibitem{hohl}J. Hohlfeld, E. Matthias, R. Knorren, and K. H. Bennemann,
Phys. Rev. Lett. {\bf 78}, 4861 (1997); {\it ibid.} {\bf 79}, 960 (1997) 
(erratum).
\bibitem{aesch} M. Aeschlimann, M. Bauer, S. Pawlik, W. Weber, 
R. Burgermeister, D.
 Oberli, and H. C.
Siegmann, Phys. Rev. Lett. {\bf  79}, 5158 (1997).
\bibitem{scholl}
A. Scholl, L. Baumgarten, and W. Eberhardt, Phys. Rev. Lett. {\bf
79}, 5146 (1997).
\bibitem{meier}
A. Vaterlaus, T. Beutler, and F. Meier, Phys. Rev. Lett. {\bf 67},
3314 (1991).
\bibitem{hz1}W. H\"ubner and G. P. Zhang, Phys. Rev. B {\bf  58}, R5920 (1998).
\bibitem{hz2}W. H\"ubner and G. P. Zhang, J. Magn. Magn. Mater. {\bf 189}, 101
(1998). 
\bibitem{fulde}P. Fulde, {\it Electron Correlation in Molecules and Solids, 3rd
edit.} (Springer, Heidelberg, 1995); J. Wahle, N. Bl\"umer, J. Schling,
K. Held, and D. Vollhardt, cond-mat/9711242.
\bibitem{hf}W. H\"ubner and L. M. Falicov, Phys. Rev. B {\bf 47}, 8783
(1993);  C. Moore, {\it Atomic Energy Levels,} Natl. Bur. Stand. (U.S.)
(U.S. GPO, Washington DC, 1971).  
\bibitem{wh}W. H\"ubner, Phys. Rev. B {\bf 42}, 11 553 (1990). 
\end{references}
\end{document}